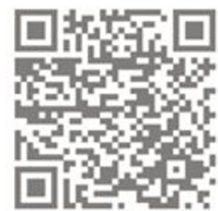

The Electrochemical Society
Advancing solid state & electrochemical science & technology

# A High Throughput Aqueous Passivation Testing Methodology for Compositionally Complex Alloys Using a Scanning Droplet Cell

View the article online for updates and enhancements.

## You may also like

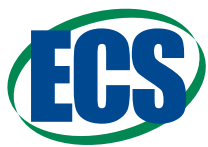

# A High Throughput Aqueous Passivation Testing Methodology for Compositionally Complex Alloys Using a Scanning Droplet Cell

Debashish Sur,[1,2,*,z] Howie Joress,[3] Jason Hattrick-Simpers,[3,4] and John R. Scully[1,2,**,z]

[1]Center for Electrochemical Science and Engineering, University of Virginia, Charlottesville, VA, 22904, United States of America
[2]Department of Materials Science and Engineering, University of Virginia, Charlottesville, VA, 22904, United States of America
[3]Materials Measurement Science Division, National Institute of Standard and Technology, Gaithersburg, MD 20899, United States of America
[4]Department of Materials Science and Engineering, University of Toronto, Toronto, ON M5S 3E4, Canada

Compositionally complex alloys containing four or more principal elements provide an opportunity to explore a wide range of compositions, processing, and microstructural variables to find new materials with unique properties. In particular, the discovery of novel alloys that form self-healing, protective passivating films is of substantial interest. Probing experimentally a robust landscape of such alloys requires the utilization of high-throughput electrochemical methods to uncover key differences, ideally captured by discriminating metrics, indicative of superior performance. Herein, a methodology is demonstrated using a scanning droplet cell for a rapid passivation behavior evaluation of $Al_{0.7-x-y}Co_xCr_yFe_{0.15}Ni_{0.15}$ combinatorial alloy library in 0.1 mol $l^{-1}$ $H_2SO_{4(aq)}$.

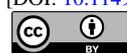

Complex concentrated alloys (CCAs) also known as Multi-principal Element Alloys and High Entropy Alloys, are metallic materials comprising four or more elemental constituents. CCAs have gained significant interest in recent years due to their ability to form a single phase and demonstrate tunable properties.[1] However, the large number of components in these alloy systems creates a vast hyperdimensional space of alloy compositions to be explored. For example, a single 5-component CCA system can have more than 0.81 million possible combinations when considering just a coarse 5 at% change in composition over a systematic range.[2,3] To explore even a single material property, such as corrosion behavior, in a space containing this many alloys demands creating workflows consisting of combining high throughput (HTp) computational and experimental techniques as well as interpretation methods to guide the screening alloys of interest.[4]

HTp techniques have been successfully applied to several topics in materials science, such as the search and discovery of desired mechanical,[5] electrical, magnetic, and optical properties.[6,7] In particular, HTp electrochemical corrosion has gained focus over the past decade as well.[8] For example, searching for chromate-free inhibitors; White et al. demonstrated a technique that can test up to 88 inhibitor solutions in an array simultaneously and in agreement with standard immersion tests.[9] Another study created a set of corrosion inhibition data with enough chemical diversity to be used to develop machine learning algorithms which can be used to predict preferred structures of small organic molecules as chromate replacements.[10] Multi-electrode arrays have been another successful rapid testing approach in delineating localized corrosion events[11,12] and testing inhibitor properties. Another rapid testing method was introduced by Kallip et al.[13] for calculating corrosion inhibition efficiencies. They used a wire beam multi-electrode cell[14] and a scanning vibrating electrode technique that allowed estimation of anodic and cathodic currents for each combination of metal and inhibitor, tested in an automated fashion.[13,15]

Scanning droplet cell (SDC) systems have come up over the last two decades as a powerful HTp electrochemical testing technique also used for aqueous corrosion research. The first attempts were aimed to miniaturize the macroscopic three-electrode arrangement of conventional electrochemical cells using capillary based microcells and stationary electrolytes.[16,17] Several applications focused on the pitting behavior of corrosion resistant materials in salt containing environments, taking advantage of an SDC's absence of any crevices, unlike in conventional cells created by O-rings.[18,19] In these cases, the primary benefit of the method was to avoid crevice occluded cell formation and capillary thin film formation which enabled differential aeration concentration cells.[19] These droplet cell designs can utilize modern additive manufacturing methods for designing crevice-free cells approaching micro-capillary cells with dimensions ranging from micrometer scale to a few mm, alleviating the constraint of testing isolated microstructural and compositional heterogeneities or their surrogates at macro scale length scales.

In recent years, SDC systems have been developed with the capability of flowing electrolytes allowing for automated testing. Fushimi et al. used such a system on brazed Al alloys, demonstrating that conventional electrochemical cells vs. SDC systems provided similar measurements of the open circuit potential (OCP) and the ability to test the contributions of Zn, Si, and Cu after brazing by testing individual phases of the microstructure.[20] Joress et al. demonstrated rapid corrosion measurements of a Zn-Ni alloy in a series of pH's aimed at coating applications.[21,22] An adaptation of a SDC has also shown potential for studying the behavior of battery electrodes such as $LiPO_4/Li$ in both aqueous and non-aqueous media in an automated fashion with the aim of designing high energy storage batteries.[23–25] More generally, a SDC can be an effective HTp tool for studying corrosion properties such as passive current densities, localized pitting potentials, and catalytic efficiency of a library of small alloy samples, enabling a data driven based search of materials and their desired attributes in the domain of electrochemistry.

A. Hassel et al., have shown a prolific amount of research on developing flowing SDC setups along with allied microscopy for HTp electrochemical evaluations.[26–28] The method was demonstrated using deposited thin film libraries containing Hf, Nb, or Ta, refractory elements[29,30] to map the electrical properties of their mixed oxides. Several other modifications were also reported that enabled microscopy at smaller length scales recording higher fidelity information *in operando*.[31–33] Another such connection was made by Gregoire et al., using a customized SDC to perform a figure of merit assessment by mapping to screen a library of $(Fe-Co-Ni-Ti)O_x$ electrocatalysts for oxygen evolution reaction.[34] These methods can be utilized to establish a workflow for screening corrosion resistant CCAs following their aqueous passivation behavior and related microstructural attributes.

*Electrochemical society member.
**Electrochemical society fellow.
zE-mail: ds8vw@virginia.edu; jrs8d@virginia.edu

In this report, we report on an HTp electrochemical rapid testing strategy and methodology designed for testing CCAs to discover indicators for good "passive" corrosion behavior of the oxide film using electrochemical impedance spectroscopy (EIS) and linear sweep voltammetry (LSV). The objective is to present a down-select methodology and useful metrics for lightweight Cr and Al containing CCAs that might suffice for further detailed consideration of alloys comparable to or exceeding stainless steels. The method will be demonstrated by using an automated flowing scanning droplet cell to characterize $Al_{0.7-x-y}Co_xCr_yFe_{0.15}Ni_{0.15}$ CCA thin films deposited as a combinatorial thin film library. Nine compositions with different concentrations of Al and Cr passivating components were chosen to demonstrate the methodology. Bulk stainless steels and an equiatomic CoCrFeMnNi[35] served as control alloys with known corrosion properties to validate the method in comparison to a conventional three electrode cell.

## Experimental

A continuous thin film library ($\approx 1\ \mu m$ thick) with varying compositions of $Al_{0.7-x}Co_xCr_yFe_{0.15}Ni_{0.15}$ was magnetron sputtered on a 3-inch wafer of $Si/SiO_2$ in a combinatorial fashion using a three-gun chamber and then annealed at 400 °C in vacuum for 9 h. The wafer was mapped and diced into 177 pieces of x-y each of area $\approx 0.2\ cm^2$. The compositional variation across a single wafer piece was less than 2 at%. High throughput (HTp) X-ray fluorescence (XRF) and X-ray diffraction (XRD) were used to obtain compositions and perform phase analysis, respectively. More details on HTp synthesis and characterization steps can be found elsewhere.[36] Nine alloys were selected for this work. Bulk homogenized coupons of commercial 316 L ($Fe_{0.69}Ni_{0.10}Cr_{0.20}Mo_{0.01}$), 304 L ($Fe_{0.73}Ni_{0.7}Cr_{0.20}$), as well as the Cantor (equiatomic CoCrFeMnNi) alloy, were also used, which were polished with #1200 grit SiC paper. All samples were ultrasonically cleaned in ethanol, and ultrapure water for 5 min and dried with $N_2$ gas.

All HTp electrochemical experiments were performed using a BioLogic Scanning Droplet Cell (SDC)[a] integrated system consisting of a SP-300 potentiostat with ultra-low current cables and M470 scanning electrochemical workstation equipped with x-y-z linear displacement actuators. The scanning head movement, as well as the electrochemical experiments, were monitored in an automated fashion using the BioLogic M470 software package. A schematic of the setup and the electrochemical sequence is shown in Fig. 1. The flow of solution was maintained using a two-channel peristaltic pump via PTFE tubes in and out of the scanning head. Before droplet formation at the scanning head opening, fresh solution travels through the reference electrode cavity. The solution then passes through the counter electrode cavity before leaving the scanning head as a waste solution, completing the electrical circuit. Optimizing the outflow and inflow rate in and out of the cavity enabled a droplet (crevice free) to be stabilized. A miniaturized saturated Ag/AgCl (E = +0.197 V vs. SHE) was used as the reference electrode, a Pt wire within the capillary as counter electrode ( area = 0.3 cm²), and the alloy thin film sample was used as the working electrode. The aperture of the scanning head was 0.2 cm wide, while the exposure area of the droplet was 0.3 cm in diameter. Test solutions of 0.1 mol l⁻¹ $H_2SO_{4(aq)}$ were prepared using reagent grade sulfuric acid (Fisher Scientific) and ultrapure water (Millipore Sigma, 18.2 $M\Omega \cdot cm$).

For performing the HTp SDC experiment, the wafer was diced into square chips (4.5 mm × 4.5 mm) and the measured 9 samples were positioned as an array in a single row on a conducting Al plate simulating their orientation on the library wafer grid (refer Fig. 2a). Silver paint was used to connect the alloy film to the backside of the wafer to enable a conductive path to the Al plate and subsequently the potentiostat. The inflow and outflow rates were fixed at 200 $\mu l\ min^{-1}$ and 400 $\mu l\ min^{-1}$, respectively. The distance between the aperture and the thin film was maintained between approximately 50 $\mu m$ to 100 $\mu m$.

All alloys using the SDC were tested following the electrochemical sequence shown in Fig. 1. After finishing the sequence for one alloy the scanning head automatically moved to another alloy position such that a 1 min delay was programmed for this action. This allowed an exchange of fresh solution and the formation of a new droplet, minimizing any contamination. For each alloy, a 10 s delay was set to stabilize the newly formed droplet before initiating the HTp sequence. The open circuit potential (OCP) was monitored for 15 min. After which, potentiostatic EIS was performed at OCP using a 20 mV (RMS) AC signal, across a frequency range of 50 kHz to 10 mHz, recording 5 points per decade. This step interrogated the corrosion reactivity of the electrolyte exposure modified aged passive films. Post EIS, a cathodic reduction step was performed at −0.76 V vs. SHE for 10 s to minimize the presence of this film. The OCP was observed for 5 s as a reference to perform an LSV from −0.3 V vs. OCP to +1.2 V vs. SHE at a scan rate of 5 mV s⁻¹. The fast scan was necessitated to minimize a potentially large magnitude of anode charge such that the electrode was not dissolved completely. This step examined the passivation of the alloy while a fresh passivating film is formed in the electrolyte. The total elapsed time per alloy was ∼26 min. A Python script was used to analyze the generated data files to extract the quantitative performance metrics in a high throughput fashion, the script can be found here.[37]

A comparison of electrochemical corrosion behavior characterization methods was performed for two polished bulk alloy samples of 316 L and Cantor by comparing corrosion assays obtained using our HTp SDC approach with that of a conventional large three electrode cell (volume of 300 ml). For the conventional cell, a $Hg/Hg_2SO_4$ (E = +0.640 V vs. SHE) reference electrode was used along with a Pt mesh counter electrode. The alloy was the working electrode. A 0.29 cm wide Viton O-ring sealed the exposed area, a size comparable to that of SDC. The electrolyte and potentiostat used and the HTp sequence followed were kept the same as mentioned above. The solution was deaerated with $N_2$ gas throughout the experiments.

## Results

The deposited combinatorial CCA $Al_{0.7-x-y}Co_xCr_yFe_{0.15}Ni_{0.15}$ thin film wafer grid library with 177 alloys with their elemental alloy compositions of {Al+Cr} obtained from XRF are shown as a color map across the wafer grid in Fig. 2a. XRD patterns of the selected 9 alloys are shown in Fig. 2b. Since Al compositions change from only 21 at% to 19 at%, hereafter the alloy set would be designated with their {Al + Cr} concentration 'y' as well. Based on our previous work on this library, alloys were classified into "*BCC*"[b] and "*Multiphase*" categories.[36] A trend of changing crystal structure from FCC (face centered cubic) + BCC (body centered cubic), termed as "*multiphase*," to single phase BCC was observed with an increase in y. This can be attributed to decreasing Co content, a known FCC stabilizer while increasing that of Cr across the row, a known BCC stabilizer, with minimal changes to Al, Fe, and Ni.

Electrochemical corrosion studies were conducted using the automated SDC. Figure 3a shows a comparison of the anodic LSV behavior for the Cantor and 316 L bulk homogenized alloys, measured following the same sequence of electrochemical experiments shown in Fig. 1 but conducted using the conventional three electrode flat cell compared to the SDC. A minimal difference in results can be observed with similar corrosion potentials ($E_{corr}$), critical potential ($E_{crit}$), critical ($i_{crit}$), and passive ($i_{pass}$) current densities independent of the cell types.[38]

[b]The "BCC" designation would represent an alloy based on its X-ray diffraction pattern having BCC as the majority phase with second phases like B2, FCC being negligible in phase fraction due to their extremely low intensity or unclear peaks.

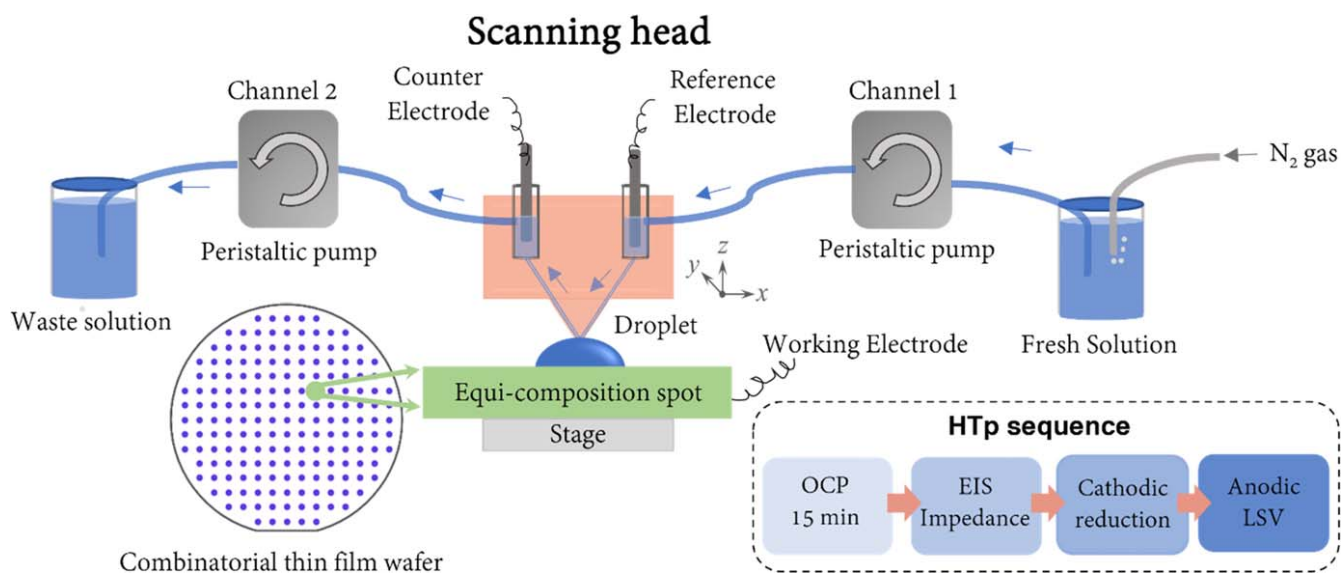

**Figure 1.** A schematic showing the high throughput scanning droplet cell setup used for this work. An Ag/AgCl (satd. KCl) and a thin Pt wire sealed inside in their sealed cavities were used as reference and counter electrodes, respectively. The arrows (blue) represent the flow direction of the test solution. The test alloy (green) represents one of the 177 alloy regions of the wafer (blue dots) that is undergoing testing.

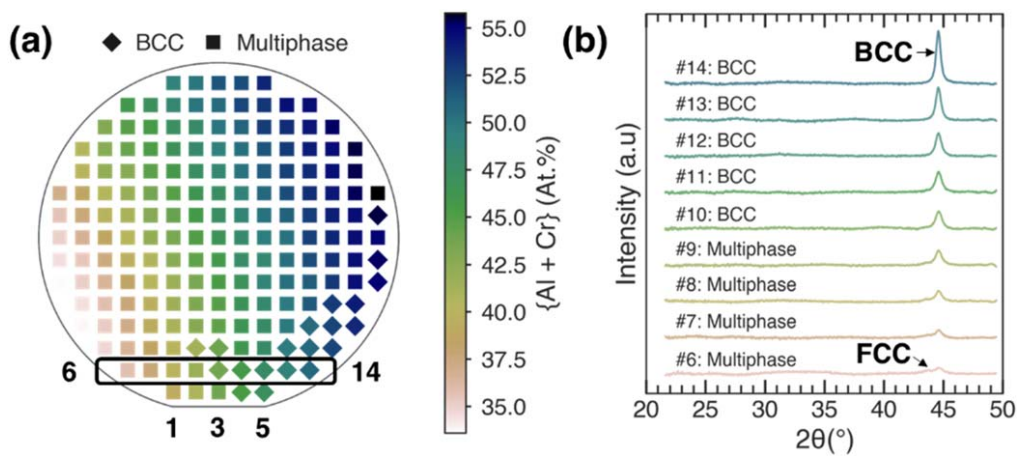

**Figure 2.** (a) x-y coordinate wafer grid of all 177 alloys, where the equi-composition alloy regions are numbered 1 to 177 in a raster fashion as labeled. The color map indicates the {Al + Cr} composition of each alloy obtained using HTp XRF. The 9 selected alloys are shown inside the encircled black rectangle numbered #6 through #14. (b) HTp XRD patterns of the nine alloys were obtained using a Cu K$\alpha$ target.

Figure 3b shows a Bode plot of the EIS at OCP for all the tests on the thin film and bulk alloys after exposing their native oxides to 0.1 mol l$^{-1}$ H$_2$SO$_{4(aq)}$ for 15 min. Here, the impedance is mainly capacitive in the case of 316 L, 304 L, #13, and #14. Other materials are active and EIS data indicates the overall interfacial electro-dissolution rate for the alloying elements below $E_{crit}$ and well above the oxidation half-cell potential of each element (Fig. 3c). A trend was observed of increasing |Z| for all the alloys with increasing y at a frequency of 10 mHz. This indicates its utility as a meaningful metric providing some description of the corrosion behavior of each alloy shown in Fig. 3d. Figure 3c presents the upward LSV results after performing the cathodic reduction step which probes the short-term passivation behavior of the bare alloy in ~5 min. The analysis of the parameters |Z|, $E_{corr}$, $i_{pass}$, and $i_{crit}$ achieved using the Python script[37] of the raw AC and DC data is shown in Table I. Almost all tested alloys showed an active-passive transition as expected in a strongly acidic environment,[39] with similar $E_{corr}$. Alloys with atom fractions y = {0.36, 0.38, 0.40, 0.42} displayed a distinctively different behavior with higher $i_{pass}$ and $i_{crit}$ and lower |Z| compared to the other alloys as shown in Fig. 3d. We attribute this to their multiphase microstructures consisting of BCC and FCC (Al depleted) phases due to large Co content as well as the poor passivation qualities of cobalt (II) itself.[40]

Another trend was observed with decreasing $i_{pass}$ and $i_{crit}$ within alloys with atom fractions y = {0.44, 0.46, 0.48, 0.49, 0.51} following Fig. 3d. All these alloys have a single phase microstructure. Such behavior could be attributed to their increasing total concentration of passivating components in the alloy. The increase in {Al + Cr} is important for causing short range ordering of alloying elements in a way that contributes to passivity.[41] Further, this composition variation was reflected in the compositions of their passive films with higher Al$_2$O$_3$ and Cr$_2$O$_3$ content correlated with improved corrosion properties.[36] The best alloys possessed atom fractions y = {0.49 and 0.51} These alloys were not as good as 316 L stainless steel but were better than the Cantor alloy based on their |Z| and $i_{crit}$ values. For alloys with single phase microstructures, |Z| and y were observed to be good indicators of low $i_{pass}$, low $i_{crit}$, and were superior to $E_{corr}$ as indicators.

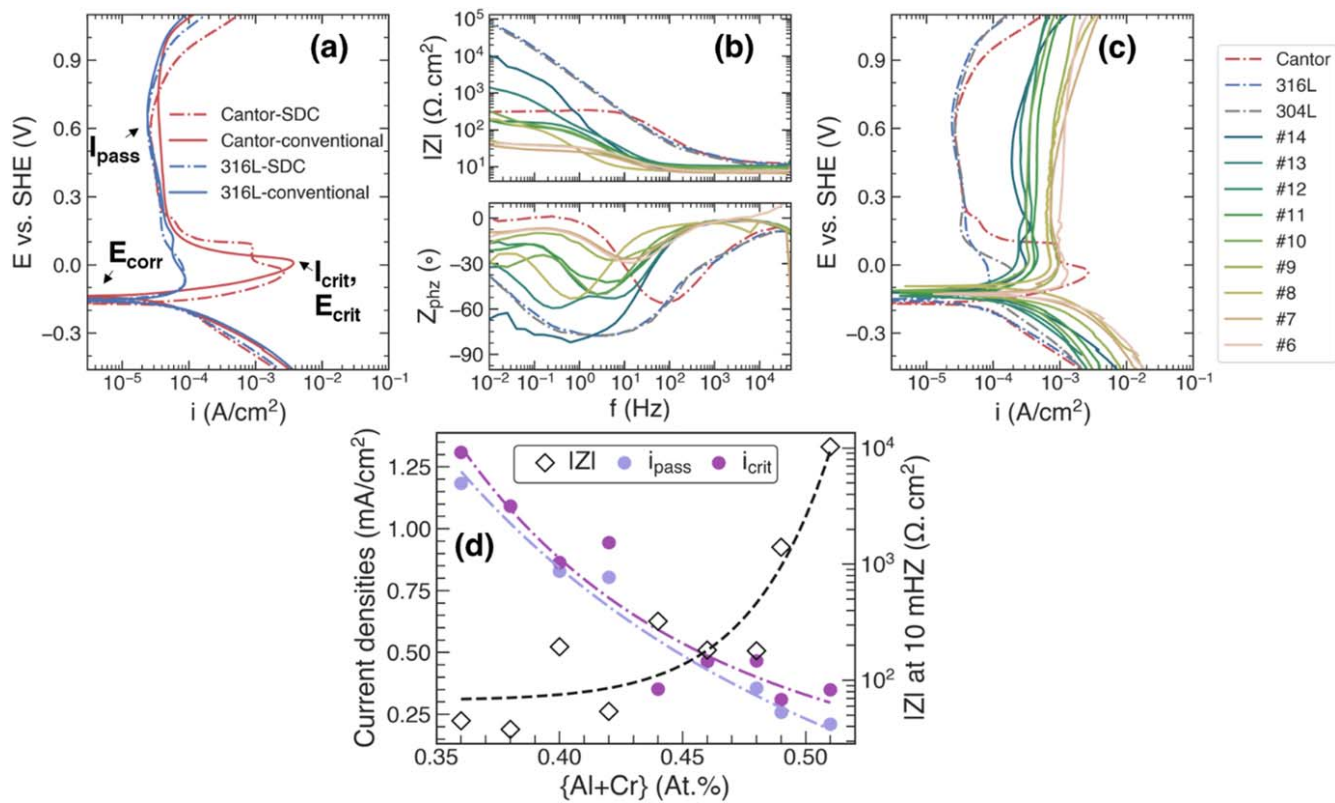

**Figure 3.** (a) A summary of comparative anodic LSV behavior of 316 L stainless steel and Cantor alloy (CoCrFeMnNi) obtained using the conventional cell and the HTp SDC in deaerated 0.1 mol l$^{-1}$ H$_2$SO$_{4(aq)}$, both following the same procedure shown in Fig. 1. (b) HTp SDC EIS Bode plots of the exposure aged native oxide passive films of thin film CCAs and bulk alloys after 15 min at OCP (c) HTp SDC LSV performed at an anodic scan rate of 5 mV s$^{-1}$ after cathodic reduction step at −0.76 V vs. SHE for 10 s, both in deaerated 0.1 mol l$^{-1}$ H$_2$SO$_{4(aq)}$. The color scheme representing the {Al+Cr} contents first established in Fig. 2, is followed here for the thin film CCAs. (d) Correlation plot of analyzed passivation metrics as a function of {Al+Cr} concentrations showing the benefits of higher {Al+Cr} concentrations towards i$_{pass}$ and |Z|.

**Table I.** Electrochemical metrics as obtained from high throughput SDC testing using the Python script.[37] Parameters include |Z| at 10 mHz from EIS on the exposure aged native oxide at OCP and E$_{corr}$ (V vs. SHE), i$_{pass}$ (at +0.6 V vs. SHE), and i$_{crit}$ from LSV after cathodic reduction. As a high-throughput measurement, the systematic trend in these measurements generally outweighs the calculable measurement uncertainties. Based on our experience, we estimate the uncertainty (2 standard deviations) in E$_{corr}$ to be ± 0.05 V and ± 0.01 mA cm$^{-2}$ for i$_{pass}$ and i$_{crit}$. For |Z| we estimate it to be ± 10 Ω·cm², and ± 100 Ω·cm² for values smaller and larger than 1000 Ω·cm², respectively. We estimate the uncertainty in the composition to be ± 0.02.

| Alloy name | Composition | $y$ | |Z| (Ω·cm²) | E$_{corr}$ (V) | i$_{pass}$ (mA cm$^{-2}$) | i$_{crit}$ (mA cm$^{-2}$) |
|---|---|---|---|---|---|---|
| #6 | Al$_{0.20}$Co$_{0.34}$Cr$_{0.16}$Fe$_{0.15}$Ni$_{0.15}$ | 0.36 | 45 | −0.13 | 1.18 | 1.31 |
| #7 | Al$_{0.20}$Co$_{0.32}$Cr$_{0.18}$Fe$_{0.15}$Ni$_{0.15}$ | 0.38 | 38 | −0.12 | 1.09 | 1.09 |
| #8 | Al$_{0.21}$Co$_{0.30}$Cr$_{0.19}$Fe$_{0.15}$Ni$_{0.15}$ | 0.40 | 195 | −0.09 | 0.83 | 0.86 |
| #9 | Al$_{0.21}$Co$_{0.28}$Cr$_{0.21}$Fe$_{0.15}$Ni$_{0.15}$ | 0.42 | 54 | −0.14 | 0.80 | 0.94 |
| #10 | Al$_{0.20}$Co$_{0.26}$Cr$_{0.24}$Fe$_{0.15}$Ni$_{0.15}$ | 0.44 | 323 | −0.11 | 0.36 | 0.35 |
| #11 | Al$_{0.20}$Co$_{0.24}$Cr$_{0.26}$Fe$_{0.15}$Ni$_{0.15}$ | 0.46 | 182 | −0.12 | 0.46 | 0.46 |
| #12 | Al$_{0.20}$Co$_{0.22}$Cr$_{0.28}$Fe$_{0.15}$Ni$_{0.15}$ | 0.48 | 180 | −0.13 | 0.36 | 0.47 |
| #13 | Al$_{0.19}$Co$_{0.21}$Cr$_{0.30}$Fe$_{0.15}$Ni$_{0.15}$ | 0.49 | 1400 | −0.12 | 0.26 | 0.31 |
| #14 | Al$_{0.19}$Co$_{0.19}$Cr$_{0.32}$Fe$_{0.15}$Ni$_{0.15}$ | 0.51 | 10251 | −0.12 | 0.21 | 0.35 |
| Cantor | Co$_{0.2}$Cr$_{0.2}$Fe$_{0.2}$Mn$_{0.2}$Ni$_{0.2}$ | – | 297 | −0.17 | 0.03 | 2.68 |
| 304 L | Fe$_{0.73}$Ni$_{0.07}$Cr$_{0.20}$ | – | 81966 | −0.12 | 0.03 | 0.21 |
| 316 L | Fe$_{0.69}$Ni$_{0.10}$Cr$_{0.20}$Mo$_{0.01}$ | – | 74045 | −0.15 | 0.02 | 0.09 |

## Discussion

The SDC electrochemical procedure as shown in Fig. 1 enabled probing of two essential performance features of aqueous passivation in an HTp fashion. One, native oxide film[42,43] resistance to corrosion after exposure to strong acid using EIS and, second, the ability of the bare alloy to passivate vis-à-vis self-healing processes, following the active and passive regions of LSV.[39]

Regarding the first, studies have shown corrosion resistant alloys with good passivation attributes undergo exposure aging to produce higher impedance after long times in chloride-free sulfate as well as chloride-containing environments. This is indicative of film growth, composition enhancement, and or annealing of defects or both, while those poorly passivating, exposure age to produce an inferior impedance over time, indicative of oxide thinning, composition change, morphology roughening, and/or defect injection. Fifteen minutes is short but may be meaningful given a 2–4 nm thick passive film. These exact details can be elucidated in follow-on high fidelity studies which can address many of these questions.[42,44–49] Moreover, a similar HTp approach can be developed in sulfate and chloride with a better emphasis on chloride breakdown. Some indicative metrics might differ while others such as self-healing are likely of similar value toward guiding the discovery of improved alloys. Further, following new approaches of HTp equivalent circuit model fitting of EIS data, more mechanisms and interfacial information can be extracted.[50]

While the differences in measurement parameters and associated measurement time lead to some variation from conventionally reported values,[51,52] we posit that, for a quick comparison between alloy compositions, the accuracy of our measurements is sufficient to determine critical trends in acidified sulfate environment. For HTp SDC testing approach, the relevant parameters of a high fidelity sequence can be customized for a given set of experimental conditions, including electrolyte, pH, cathodic reduction potential and hold time, flow rates, and droplet dimensions. For instance, pitting resistance requires halide containing solutions and different metrics. In the case of the alloys discussed herein, high fidelity studies confirmed that down selects made in this study based on the limited information herein were good choices in subsequent exposures and higher fidelity electrochemical testing.[36] High Cr and Al translate into superior oxide and protect fullness even at low Cr levels relative to classical critical Cr levels. The addition of Al, a lower density element showed promise in substituting for some of the Cr content but can form secondary phases detrimental to corrosion resistance.[53] Thus, {Al+Cr} and the single phase nature of alloys are good indicators of corrosion resistance which tracks with the parameter $y$.

Further, the thin films CCAs showed lower |Z| (Fig. 3b) and higher $i_{pass}$ (Fig. 3c) compared to bulk alloys, which can be partially attributed to pore passivity as well as their porous surfaces increasing the true exposed surface area.[36] CoCrFeMnNi showed a larger $i_{crit}$, and lower impedance compared to the thin film CCAs that could be attributed to its 20 at% Mn content, as Mn selectively dissolves with negligible participation towards alloy passivation.[44,45] Suitability of sputtered thin films as surrogates is often a question in corrosion research as far as how accurately they represent a bulk alloy produced conventionally.[54] This is the subject of our separate studies and is beyond the scope of the current manuscript.

## Conclusions

A high throughput aqueous passivation performance methodology is demonstrated using a scanning droplet cell and a combinatorial thin film for an $Al_{0.7−x−y}Co_xCr_yFe_{0.15}Ni_{0.15}$ alloy system. A series of 9 CCAs from the library were tested within a total period of 4 h of testing time, taking $\approx$26 min per alloy. Metrics such as the EIS based impedance at 10 mHz, measured on the acid exposure aged native oxide layer, and the LSV-derived passive current density for the anodically grown passive layer were used to judge the performance of selected alloys with specific oxide films. Moreover, self-healing was assessed during electrochemical passivation. Multiphase CCA metrics showed a trend with high/low performance in sulfate. Within single phase alloys, the performance was enhanced with increasing {Al + Cr} concentrations in the alloys.

## Acknowledgments

The authors gratefully acknowledge partial funding for this work from the Office of Naval Research through the Multidisciplinary University Research Initiative program (award #: N00014-20-1-2368), with program manager Dr. David Shifler. Thanks also to Dan Gopman for dicing the wafers at the NIST Center for Nanoscale Science and Technology.

## ORCID

Debashish Sur  https://orcid.org/0000-0002-7954-9949
Howie Joress  https://orcid.org/0000-0002-6552-2972
Jason Hattrick-Simpers  https://orcid.org/0000-0003-2937-3188
John R. Scully  https://orcid.org/0000-0001-5353-766X

## References

1. D. Miracle et al., *Defining Pathways for Realizing the Revolutionary Potential of High Entropy Alloys: A TMS Accelerator Study* (The Minerals, Metals & Materials Society (TMS), Pittsburgh, PA) (2021), 10.7449/HEApathways.
2. D. B. Miracle, "High entropy alloys as a bold step forward in alloy development." *Nat. Commun.*, **10**, 1805 (2019).
3. R. Feng et al., "Design of light-weight high-entropy alloys." *Entropy*, **18**, 16 (2016).
4. J. J. Bhattacharyya et al., "Lightweight, low cost compositionally complex multiphase alloys with optimized strength, ductility and corrosion resistance: Discovery, design and mechanistic understandings." *Mater. Des.*, **228**, 111831 (2023).
5. T. Oellers, V. G. Arigela, C. Kirchlechner, G. Dehm, and A. Ludwig, "Thin-film microtensile-test structures for high-throughput characterization of mechanical properties." *ACS Comb. Sci.*, **22**, 142 (2020).
6. R. A. Potyrailo and V. M. Mirsky, "Combinatorial and high-throughput development of sensing materials: The first 10 years." *Chem. Rev.*, **108**, 770 (2008).
7. H. Joress, M. L. Green, I. Takeuchi, and J. R. Hattrick-Simpers, "Applications of high throughput (combinatorial) methodologies to electronic, magnetic, structural, and energy-related materials." *Encyclopedia of Materials: Metals and Alloys* (Elsevier, Amsterdam), p. 353 (2021), 10.1016/B978-0-12-819726-4.00146-0.
8. T. H. Muster et al., "A review of high throughput and combinatorial electrochemistry." *Electrochim. Acta*, **56**, 9679 (2011).
9. P. A. White et al., "A new high-throughput method for corrosion testing." *Corros. Sci.*, **58**, 327 (2012).
10. D. A. Winkler et al., "Using high throughput experimental data and in silico models to discover alternatives to toxic chromate corrosion inhibitors." *Corros. Sci.*, **106**, 229 (2016).
11. T. T. Lunt, V. Brusamarello, J. R. Scully, and J. L. Hudson, "Interactions among localized corrosion sites investigated with electrode arrays." *Electrochem. Solid-State Lett.*, **3**, 271 (2000).
12. N. D. Budiansky, J. L. Hudson, and J. R. Scully, "Origins of persistent interaction among localized corrosion sites on stainless steel." *J. Electrochem. Soc.*, **151**, B233 (2004).
13. S. Kallip, A. C. Bastos, M. L. Zheludkevich, and M. G. S. Ferreira, "A multi-electrode cell for high-throughput SVET screening of corrosion inhibitors." *Corros. Sci.*, **52**, 3146 (2010).
14. Y. J. Tan, "Wire beam electrode: a new tool for studying localised corrosion and other heterogeneous electrochemical processes." *Corros. Sci.*, **41**, 229 (1998).
15. S. Fajardo, C. F. Glover, G. Williams, and G. S. Frankel, "The source of anodic hydrogen evolution on ultra high purity magnesium." *Electrochim. Acta*, **212**, 510 (2016).
16. T. Suter and H. Böhni, "A new microelectrochemical method to study pit initiation on stainless steels." *Electrochim. Acta*, **42**, 3275 (1997).
17. M. M. Lohrengel, A. Moehring, and M. Pilaski, "Electrochemical surface analysis with the scanning droplet cell." *Fresenius J. Anal. Chem.*, **367**, 334 (2000).
18. H. Y. Ha, C. J. Park, and H. S. Kwon, "Effects of non-metallic inclusions on the initiation of pitting corrosion in 11% Cr ferritic stainless steel examined by micro-droplet cell." *Corros. Sci.*, **49**, 1266 (2007).
19. A. M. Panindre and G. S. Frankel, "Technical note: electrochemical testing for pitting corrosion under ambient temperatures using the syringe cell." *Corrosion*, **77**, 1025 (2021).
20. K. Fushimi, S. Yamamoto, R. Ozaki, and H. Habazaki, "Cross-section corrosion-potential profiles of aluminum-alloy brazing sheets observed by the flowing electrolyte scanning-droplet-cell technique." *Electrochim. Acta*, **53**, 2529 (2008).
21. H. Joress et al., "Development of an automated millifluidic platform and data-analysis pipeline for rapid electrochemical corrosion measurements: A pH study on Zn-Ni." *Electrochim. Acta*, **428**, 140866 (2022).
22. H. Joress et al., "A High-throughput structural and electrochemical study of metallic glass formation in Ni-Ti-Al." *ACS Comb. Sci.*, **22**, 330 (2020).
23. S. Daboss, F. Rahmanian, H. S. Stein, and C. Kranz, "The potential of scanning electrochemical probe microscopy and scanning droplet cells in battery research." *Electrochem. Sci. Adv.*, **2**, e2100122 (2022).
24. G. García, S. Dieckhöfer, W. Schuhmann, and E. Ventosa, "Exceeding 6500 cycles for LiFePO4/Li metal batteries through understanding pulsed charging protocols." *J. Mater. Chem. A*, **6**, 4746 (2018).
25. S. Dieckhöfer, W. Schuhmann, and E. Ventosa, "Accelerated electrochemical investigation of Li plating efficiency as key parameter for Li metal batteries utilizing a scanning droplet cell." *ChemElectroChem*, **8**, 3143 (2021).
26. J. P. Kollender, M. Voith, S. Schneiderbauer, A. I. Mardare, and A. W. Hassel, "Highly customisable scanning droplet cell microscopes using 3D-printing." *J. Electroanal. Chem.*, **740**, 53 (2015).
27. S. O. Klemm, J. P. Kollender, and A. Walter Hassel, "Combinatorial corrosion study of the passivation of aluminium copper alloys." *Corros. Sci.*, **53**, 1 (2011).
28. J. P. Kollender, A. I. Mardare, and A. W. Hassel, "Multi-scanning droplet cell microscopy (multi-SDCM) for truly parallel high throughput electrochemical experimentation." *Electrochimica Acta*, **179**, 32 (2015).

29. A. I. Mardare, A. Ludwig, A. Savan, and A. W. Hassel, "Scanning droplet cell microscopy on a wide range hafnium-niobium thin film combinatorial library." *Electrochim. Acta*, **110**, 539 (2013).

30. A. I. Mardare, A. Ludwig, A. Savan, A. D. Wieck, and A. W. Hassel, "Combinatorial investigation of Hf-Ta thin films and their anodic oxides." *Electrochim. Acta*, **55**, 7884 (2010).

31. J. Gasiorowski, A. I. Mardare, N. S. Sariciftci, and A. W. Hassel, "Characterization of local electrochemical doping of high performance conjugated polymer for photovoltaics using scanning droplet cell microscopy." *Electrochim. Acta*, **113**, 834 (2013).

32. J. P. Kollender, J. Gasiorowski, N. S. Sariciftci, A. I. Mardare, and A. W. Hassel, "Photoelectrochemical and electrochemical characterization of sub-micro-gram amounts of organic semiconductors using scanning droplet cell microscopy." *J. Phys. Chem. C*, **118**, 16919 (2014).

33. C. M. Siket, A. I. Mardare, M. Kaltenbrunner, S. Bauer, and A. W. Hassel, "Surface patterned dielectrics by direct writing of anodic oxides using scanning droplet cell microscopy." *Electrochim. Acta*, **113**, 755 (2013).

34. J. M. Gregoire, C. Xiang, X. Liu, M. Marcin, and J. Jin, "Scanning droplet cell for high throughput electrochemical and photoelectrochemical measurements." *Rev. Sci. Instrum.*, **84**, 024102 (2013).

35. B. Cantor, "Multicomponent and high entropy alloys." *Entropy*, **16**, 4749 (2014).

36. D. Sur et al., *High Throughput Discovery of Lightweight Corrosion-Resistant Compositionally Complex Alloys* (2023).

37. D. Sur, "Electrochemical parameters." (2023), (https://github.com/debashtrix/Corrosion-science.git).

38. J. R. Scully and K. Lutton, "Polarization behavior of active-passive metals and alloys." *Encyclopedia of Interfacial Chemistry: Surface Science and Electrochemistry* (Elsevier, Amsterdam) p. 439 (2018).

39. H. H. Uhlig and G. E. Woodside, "Anodic polarization of passive and non-passive chromium–iron alloys." *J. Phys. Chem.*, **57**, 280 (1952).

40. Y. Shi et al., "Corrosion of Al CoCrFeNi high-entropy alloys: Al-content and potential scan-rate dependent pitting behavior." *Corros. Sci.*, **119**, 33 (2017).

41. E. A. Anber, N. C. Smith, and P. K. Liaw, "Role of Al additions in secondary phase formation in CoCrFeNi high entropy alloys." *APL Mater.*, **10**, 101108 (2022).

42. S. Inman, D. Sur, J. Han, K. M. Ogle, and J. R. Scully, "Corrosion behavior of a compositionally complex alloy utilizing simultaneous Al, Cr, and Ti Passivation." *Corros. Sci.*, **217**, 111 (2023).

43. L. Wang, A. Seyeux, and P. Marcus, "Ion transport mechanisms in the oxide film formed on 316L stainless steel surfaces studied by ToF-SIMS with 18 O 2 isotopic tracer." *J. Electrochem. Soc.*, **167**, 101511 (2020).

44. A. Y. Gerard et al., "Aqueous passivation of multi-principal element alloy Ni38Fe20Cr22Mn10Co10: Unexpected high Cr enrichment within the passive film." *Acta Mater.*, **198**, 121 (2020).

45. S. B. Inman et al., "Effect of Mn content on the passivation and corrosion of Al0.3Cr0.5Fe2MnxMo0.15Ni1.5Ti0.3 compositionally complex face-centered cubic alloys." *CORROSION*, **78**, 32 (2021).

46. N. Birbilis, S. Choudhary, J. R. Scully, and M. L. Taheri, "A perspective on corrosion of multi-principal element alloys." *Npj Mater. Degrad.*, **5**, 1 (2021).

47. K. Lutton et al., "Passivation of Ni-Cr and Ni-Cr-Mo alloys in low and high pH sulfate solutions." *J. Electrochem. Soc.*, **170**, 021507 (2023).

48. K. Lutton, W. H. Blades, J. R. Scully, and P. Reinke, "Influence of chloride on nanoscale electrochemical passivation processes." *J. Phys. Chem. C*, **124**, 9289 (2020).

49. P. Marcus, V. Maurice, and H. H. Strehblow, "Localized corrosion (pitting): A model of passivity breakdown including the role of the oxide layer nanostructure." *Corros. Sci.*, **50**, 2698 (2008).

50. R. Zhang, R. Black, D. Sur, P. Karimi, K. Li, B. DeCost, J. R. Scully, and J. Hattrick-Simpers, "Editors' Choice—AutoEIS: Automated Bayesian Model Selection and Analysis for Electrochemical Impedance Spectroscopy." *J. Electrochem. Soc.*, **170**, 086502 (2023).

51. H. Luo, Z. Li, A. M. Mingers, and D. Raabe, "Corrosion behavior of an equiatomic CoCrFeMnNi high-entropy alloy compared with 304 stainless steel in sulfuric acid solution." *Corros. Sci.*, **134**, 131 (2018).

52. Z. Wang, A. Seyeux, S. Zanna, V. Maurice, and P. Marcus, "Chloride-induced alterations of the passive film on 316L stainless steel and blocking effect of pre-passivation." *Electrochim. Acta*, **329**, 135159 (2020).

53. Y. F. Kao, T. D. Lee, S. K. Chen, and Y. S. Chang, "Electrochemical passive properties of AlxCoCrFeNi (x = 0, 0.25, 0.50, 1.00) alloys in sulfuric acids." *Corros. Sci.*, **52**, 1026 (2010).

54. J. Han et al., "Electrochemical stability, physical, and electronic properties of thermally pre-formed oxide compared to artificially sputtered oxide on Fe thin films in aqueous chloride." *Corros. Sci.*, **186**, 109456 (2021).